# On characterizing X-ray detectors for low-dose imaging

Kostiantyn Sakhatskyi[1,2], Ying Zhou[3], Vitalii Bartosh[1,2], Gebhard J. Matt[1,2], Jingjing Zhao[4], Sergii Yakunin[*1,2], Jinsong Huang[*3,5], Maksym V. Kovalenko[*1,2]

[1] Laboratory of Inorganic Chemistry, Department of Chemistry and Applied Biosciences, ETH Zürich, CH-8093 Zürich, Switzerland

[2] Laboratory for Thin Films and Photovoltaics, Empa – Swiss Federal Laboratories for Materials Science and Technology, CH-8600 Dübendorf, Switzerland

[3] Department of Applied Physical Sciences, University of North Carolina at Chapel Hill, Chapel Hill, NC, USA

[4] School of Physical Science and Technology, Chongqing Key Lab of Micro&Nano Structure Optoelectronics, Southwest University, Chongqing, 400715 China.

[5] Department of Chemistry, University of North Carolina at Chapel Hill, Chapel Hill, NC, USA

[*]E-mail: mvkovalenko@ethz.ch; jhuang@unc.edu; yakunins@ethz.ch

**The last decade has seen a renewed exploration of semiconductor materials for X-ray detection, foremost focusing on lead-based perovskites and other metal halides as direct-conversion materials and scintillators. However, the reported performance characteristics are often incomplete or misleading in assessing the practical utility of materials. This Perspective offers guidelines for choosing, estimating and presenting the relevant figures of merit. We also provide ready-to-used tools for calculating these figures of merit: MATLAB application, Mathcad worksheet and a website. The X-ray detectors for medical imaging are at focus for their increasing societal value and since they bring about the most stringent requirements as the image shall be acquired at as low as reasonably attainable (*i.e.* ALARA principle) dose received by the patient.**

Use-inspired materials exploration and engineering, and subsequent technological deployment immensely benefit from working with commonly agreed, utmost relevant figures of merit and protocols for their measurements as well as mandatory checklists for their reporting and, where possible, independent certification. Such consensus is well-established, for instance, for solar-cells[1], thermoelectrics[2], photodetectors[3], solid-state lighting [4], displays[5], batteries[6] and lasers[7], both at materials and device levels. In medical imaging, X-ray detectors are increasingly required to deliver acceptable image quality at ALARA dose[8]. A new wave of research reports propose diverse direct-conversion semiconductors[9,10,11,12,13] and scintillators[14,15,16] for X-ray detectors. However, the difficulty in assessing their practical utility is mounting for both expert readers and generalists. In this Perspective, we propose comprehensive characterization guidelines for low-dose X-ray imaging detectors (Fig. 1a,b), encompassing materials for indirect (Fig. 1c) and direct (Fig. 1d) X-ray sensing. We first highlight Detective Quantum Efficiency[17,18,19] (DQE, Fig. 1a, e-h) as a foundational quality metric for X-ray imaging systems. We then break down the protocol for the estimation of parameters that together yield DQE: Noise Equivalent Dose (NED, Fig. 1e,f), spatial resolution (Fig. 1e,g), detector speed (rise and fall time, Fig. 1e,h), Detection Efficiency (DE), X-ray energy at 50% absorption efficiency and Detective Quantum Efficiency index (DQEi). These parameters arise from the thorough inclusion of X-ray-matter interaction, charge-transport and light-propagation physics and device architectures. With these figures of merit, objective and application-relevant comparison of X-ray detectors can be drawn for any direct-conversion or scintillator material they comprise. The calculations described in this Perspective are additionally provided as the Mathcad worksheet, the MATLAB application (see Supplementary Data) and the webpage[20]. We also outline the good practices for the claims in the research articles, particularly in relation to different materials' development stages (Table 1).

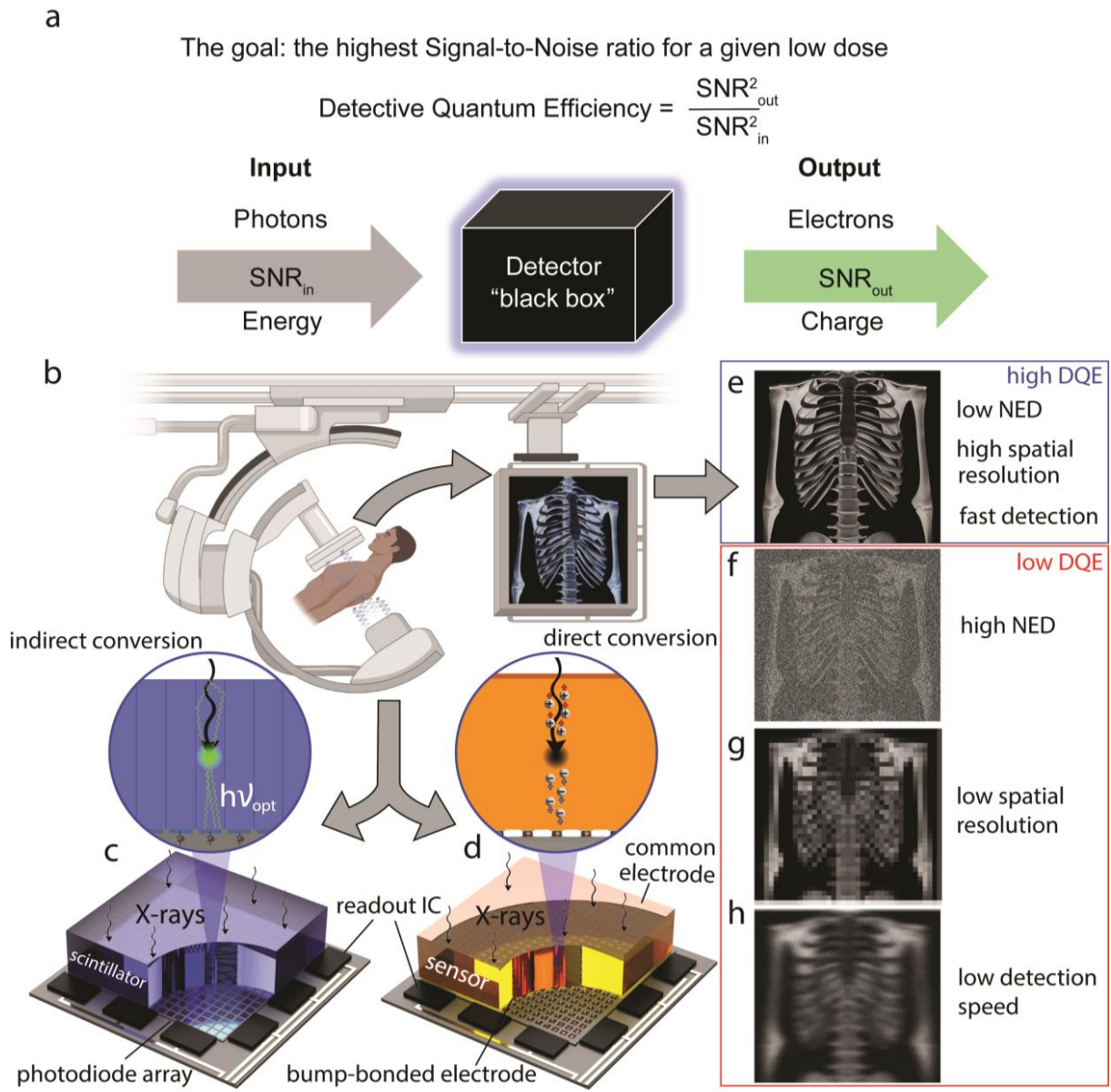

**Fig. 1. X-ray medical imaging figures-of-merits. a**, Schematics illustration of conversion processes (photons/electrons or input/output signal-to-noise ratios or energy/charge) taking place in an imaging detector independent of its type. **b** Schematic of X-ray medical imaging. **c,d**, X-ray imaging detector structure for two major X-ray-to-charge conversion technologies: indirect with scintillators (**c**) and direct with semiconductors, illustrated with the single crystal ball-bonded to the readout array (**d**), but also comprising vapor-phase-deposited a-Se commonly used in mammography application. **e**, High-quality X-ray image arises with the detector simultaneously exhibiting low NED, high spatial resolution and high speed. **f**, **g**, **h**, X-ray image quality deteriorates for either of: high noise because of high NED (**f**), blur caused by low spatial resolution (**g**), motion smearing due to low detection speed or long integration time (**h**).

**Table 1. Recommendation for reporting material and detector metrics in a research article.** Accordingly, the narrative and claims of the research article must be adjusted as to not misguide the academic, industrial and broader readership, and not raise false hopes for the fast commercial implementation (scientific populism). Overselling of the research results hampers the prospects of other, more balanced studies, to land in the same dissemination channels (journals) or receive the same credit, and distorts the assessment of the novelty and impact by the major stakeholders (funding agencies and general public).

| Stages: / Technology: | Basic material properties | Conversion factors | Detectors figures-of-merit | Commercialization feasibility |
|---|---|---|---|---|
| Direct detectors | X-ray absorption, mobility, lifetime, resistivity | X-ray sensitivity | DE, NED, $E_{1/2}$, speed, spatial resolution, DQE | Stability, integration with electronics, toxicity, prices, system-specific performance |
| Scintillators | X-ray absorption, optical Stokes shift, quantum yield | Light yield | | |
| ✓ May be claimed | X-ray detection promise | X-ray detection ability | X-ray detector performance | Commercial implementation |
| ✗ Do not claim in your paper | X-ray detection ability | X-ray detector performance | Commercial implementation | — |

For X-ray detector performance assessment, it is important to establish characterization conditions (including X-ray source collimation, energy spectrum, readout electronics, and operational temperature) commensurate to the intended application area (*e.g.* computed tomography, mammography or radiography). Even though distinct applications may require different conditions, the characterization algorithm, described below, can be applied universally for the performance comparison of diverse detectors within a designated X-ray medical imaging area.

**Single-readout-channel detectors**

Novel materials are initially tested in single-pixel devices (single-readout-channel detectors). Firstly, it is important to define the signal and the noise of the X-ray detector. A frequent mistake in the literature is considering the signal as equal to the photocurrent transient value, while the noise is calculated as its standard deviation (Fig. 2a). Those definitions of the signal and the noise do not correspond to the actual imaging process and lead to misguided evaluation of the performance, which is described in details in the section "Device temporal response characterization". We define the signal as the mean X-ray generated charge collected during a certain integration time (Fig. 2a,b). The signal should be calculated as a difference between the mean total collected charge under X-ray exposure and the mean accumulated charge in the dark at the same integration time. The noise is defined as the standard deviation of the signal Gaussian distribution, obtained after multiple signal acquisitions under the same X-ray exposure (Fig. 2b). This is notably true for both scintillation and direct-conversion detectors, as for the first case the charge is collected in optical photodetector, while, in the second case, the charge is collected directly in the sensor material. Such definition is also independent on detection mode, whether it is charge-integration or photon counting, as both an ADC unit and a photon counting event are proportional to a collected charge. Signal *vs.* dose dependence should be linear for the particular dose range, well above the noise threshold, regardless of used units (Fig. 2c):

$$\text{Signal}(D) = \text{DE} \cdot \alpha \cdot G \cdot D, \tag{1}$$

where DE is defined as the ratio of detected photons to incoming ones. DE is the integral product of X-ray absorption efficiency and collection efficiency of a generated charge (for direct detection) or light (for scintillators), which for the last case includes the scintillator form factor, reflective and refraction properties of sensor, and optical photodetector. While $\alpha$ is a conversion coefficient that describes the conversion of the dose (*i.e.* X-ray photon number) into charges (*i.e.* X-ray sensitivity for direct-conversion) or into optical photons (*i.e.* light yield for scintillators). Units of $\alpha$ are dependent on units in which the signal and the dose are expressed (3 pairs of axis in Fig. 2c.). Independent evaluation of $\alpha$ might be rather complicated, especially for a scintillator detector, as the light yield may depend on X-ray energy[21] and may be quenched at the high X-ray doses. $G$ is an electronic gain of the detector, which may include a photoconductive gain of an active material for the direct-conversion materials. For the case when Eq. 1 is applicable and the signal *vs.* dose dependence is measured, it is possible to determine the entire DE·$\alpha$·$G$ product. Estimating of each parameter separately might require a dedicated experiment. If the detector system is able to count single photons ($\alpha$ corresponds to 1 count/photon, $G$=1, case A in Box 1) or it is working in a charge-integration mode with specific $G$ (case B in Box 1), then the product DE·$\alpha$·$G$ value (obtained from Fig. 2c) may be used for direct determination of DE:

$$\text{DE} = \frac{n_{photon\_detected}}{n_{photon\_incoming}} = \frac{\text{Signal}(D)}{\alpha \cdot G \cdot D}. \tag{2}$$

For example, for a direct-conversion semiconductor detector $\alpha$ is equal to the product of detector area ($A$) and the theoretical X-ray sensitivity of photodiode, which is possible to derivate using Eq.14 and Eq.16 from Ref. 22 and assuming Klein rule for electron-hole pair creation[23]:

$$\text{X-ray sensitivity} = C \frac{e}{\left(\frac{\mu}{\rho}\right)_{air} W_\pm}, \tag{3}$$

where $e$ is the elementary charge, $W_\pm$ is electron-hole pair creation energy by ionization[23], $\left(\frac{\mu}{\rho}\right)_{air}$ – X-ray mass energy absorption coefficient in the air (X-ray energy dependent), C is the constant dependent on used units. For example, if X-ray sensitivity is expressed in units of µC cm$^{-2}$ Gy$^{-1}$$_{air}$, $W_\pm$ units are eV, $\left(\frac{\mu}{\rho}\right)_{air}$ units are cm$^2$ g$^{-1}$, then C is equal to $10^3$ µC eV g$^{-1}$ Gy$^{-1}$$_{air}$ $e^{-1}$. X-ray mass energy absorption coefficient is used specifically for the air (not for the detector sensor material!), since most dosimeters measure air kerma, which is equivalent to the absorbed dose in air for X-ray energies relevant to X-ray medical imaging applications. Also, formula (3) assumes monoenergetic X-ray photons. In the general case of a polyenergetic photon flux, X-ray sensitivity has to be averaged over the spectrum considering the dependence of $\left(\frac{\mu}{\rho}\right)_{air}$ on X-ray energy. For materials exhibiting near-100% charge collection efficiency, for instance, CdTe with the bandgap of 1.5 eV, the ultimate X-ray sensitivities, according to equation (3), would approach 3500 µC·cm$^{-2}$·Gy$^{-1}_{air}$ for 40 keV X-ray energy. Higher sensitivities in the literature are obtained due to inconsideration of $G$. In the case of difficult-to-estimate $G$ or $\alpha$, the DE is better determined from the noise analysis, as described further. Important to note, it might be complicated to evaluate the theoretical X-ray sensitivity of novel semiconductors using Eq. 3, since while $W_{\pm}$ parameter follows Klein rule[23] for most semiconductors, there are several exception. Thus, Eq. 3 should be used with cautions when $W_{\pm}$ is unknown.

The dose can be described as the energy absorbed per mass unit (J·kg$^{-1}$ or Gray) as well as a number of X-ray photons incident to the detector area ($n_{ph}$). For the homogenous parallel X-ray beam with the direction perpendicular to the detector area the dose and $n_{ph}$ are related as (it can be derived from Eq. 3.17 in Ref 24):

$$D\,[Gy_{air}] = \frac{\left(\frac{\mu}{\rho}\right)_{air} E_{ph}\, n_{ph}}{A}, \tag{4}$$

where $E_{ph}$ – X-ray photon energy. It should be noted that this formula is applicable only for a mono-energetic or quasi mono-energetic (filtered to a narrow band spectrum) photon flux incident normal to the detector surface. In a more general case, an X-ray source may have a broad energy spectrum. Thus, an averaging of $\left(\frac{\mu}{\rho}\right)_{air} \cdot E_{ph}$ product over the X-ray source spectrum is necessary. Eq. 4 is applied only for the absorbed dose in air (not for the dose in the sensor material!), since for its derivation the charge particle equilibrium is assumed - the dose build-up region is ignored and all secondary photons escape the interaction volume. Eq. 4 is still of practical value, since radiation diagnostics dosimeters commonly measure air kerma.

Noise is a root-mean-square or standard deviation of the mean signal value. Square of total noise (Noise$_T^2$) is linearly dependent (Fig. 2d) on a photon number $n_{ph}$ according to (for simplicity, $\alpha$ is expressed as a.u. per photon):

$$\text{Noise}_T^2[(a.u.)^2] = DE \cdot G^2 \cdot \alpha^2 \cdot n_{ph} + \text{Noise}_D^2. \tag{5}$$

We should note, that the described approach isn't applicable if at least one of the dependencies, equation (1) and equation (5), are substantially nonlinear. According to one definition[25], NED is the radiation dose that generates the quantum noise, that corresponds to the situation when the two right terms of equation (5) are equal. The first term is described by Poisson statistics of absorbed X-ray photons. The second term is the intrinsic noise of the detector ($Noise_D$) in the absence of radiation. Thus, NED can be determined as an offset/slope ratio from the linear fit of the experimentally measured dependency $N_T(D)$ in units of either photon number or dose:

$$\text{NED[photon]} = \frac{\text{Noise}_D^2[(\text{a.u.})^2]}{\text{DE} \cdot G^2 \cdot \alpha^2}, \qquad (6)$$

$$\text{NED[Gy}_\text{air}\text{]} = \frac{\left(\frac{\mu}{\rho}\right)_\text{air} \cdot E_{ph} \cdot \text{NED[photon]}}{A}. \qquad (6a)$$

For determination of NED, it suffices to use the experimentally obtained product $DE \cdot G^2 \cdot \alpha^2$, while estimation of separate values for DE, $G$ and $\alpha$ isn't necessary. We should note, that $Noise_D$ drastically increases with $G$, thus NED doesn't not decrease with increasing $\underline{G}$. NED in X-ray photon number units (equation (6)) points to the detector system's ability to count single photons, when NED << 1 photon. At the same time, NED in dose units (equation (6a)) is useful to report, as it determines low-dose performance. As follows from equation (6a), NED value is inversely proportional to detector area and might become attractively low for large detectors. While it is appealing for dosimetry, it is, however, not beneficial for medical imaging since it harms spatial resolution. It is therefore crucial, for medical imaging, that both parameters of NED[Gy$_\text{air}$] and spatial resolution or at least pitch size, when spatial resolution is unknown, are reported simultaneously.

To determine DE directly from the noise analysis, the dependence $Noise_T$ on dose should be represented in photon-equivalents - a factor that scales the noise amplitude to the signal of a single incident X-ray photon. The equation (5) should thus be divided by the product $(DE \cdot G \cdot \alpha)^2$, known from the equation (1) (Fig. 2c):

$$\text{Noise}_T^2[(\text{photon-equvialent})^2] = \frac{n_{\text{ph}}}{\text{DE}} + \text{Noise}_D^2. \qquad (7)$$

DE can be extracted as the inverse slope parameter from the linear function fit of noise values (Fig. 2d), according to equation (7), (case C in Box 1).

NED and DE can be combined into a single characteristic – DQE (Fig. 2e). The latter is often formulated[26] as a squared ratio of the output detector signal-to-noise, $\text{SNR}_{\text{out}}^2 = \frac{(DE \cdot \alpha \cdot G \cdot n_{ph})^2}{DE \cdot G^2 \cdot \alpha^2 \cdot n_{ph} + \text{Noise}_D^2}$, as obtained from equations (1) and (5), to the input (or ideal) signal-to-noise ratio, $\text{SNR}_{\text{in}}^2 = n_{ph}$, defined by Poisson statistics of the incident X-ray photon flux[27]. Taking also into account equations (6, 6a), one obtains:

$$\text{DQE}(D) = \frac{\text{SNR}_{\text{out}}^2}{\text{SNR}_{\text{in}}^2} = \frac{\text{DE}}{1 + \frac{\text{NED[photon]}}{n_{ph}}} = \frac{\text{DE}}{1 + \frac{\text{NED[Gy]}}{D}}. \tag{8}$$

DQE characterizes the square of the detector signal-to-noise ratio gained per unit of photons quantity, i.e., exposure dose. DQE is commonly studied as the function dependent on the spatial frequency[18] for array detectors, but is a valid characteristic to report also for single-pixel detectors.

Typically, an X-ray tube with polyenergetic spectrum is used for X-ray medical imaging. Such spectrum is characterized by standard diagnostic beam quality series[28]. For the X-ray detection performance characterization equivalently valid to use X-ray spectrum with specific beam quality (close to intended application) or to apply monochromatic or narrow width spectrum with a mean X-ray photon energy ($E_{\text{mes}}$), which could be obtained with set of attenuators[29], monochromators[30] or by using radioactive gamma-emitting isotopes. The (quasi)monochromatic X-ray spectrum can be particularly beneficial for the detector characterization, since it allows to evaluate the entire operational X-ray energy range by extrapolating DQE ($n_{ph}$), NED and DE (Fig. 2f) using the dispersion of X-ray Absorption Efficiency (AE), (case D in Box 1):

$$\text{AE}(E_{ph}) = \left(1 - \exp(-\mu(E_{ph})d)\right), \tag{9}$$

$$\text{NED}(E_{ph}) \approx \frac{\text{NED}(E_{mes}) \times E_{mes} \times \text{AE}(E_{ph})}{E_{ph} \times \text{AE}(E_{mes})}, \tag{10}$$

$$\text{DE}(E_{ph}) \approx \text{DE}(E_{mes}) \frac{\text{AE}(E_{ph})}{\text{AE}(E_{mes})}, \tag{11}$$

$$\text{DQE}(E_{ph}, n_{ph}) \approx \frac{\text{DE}(E_{ph})}{1 + \frac{\text{NED}(E_{ph})}{n_{ph}}}. \tag{12}$$

Where $\mu$ is the linear absorption coefficient (dependent on $E_{ph}$, which can be calculated using the atomic composition and specific mass[31]), $d$ is the detector thickness. The plot of energy-dependent DQE for different photon numbers shows that at each dose there is optimal energy range. At high doses, DQE becomes limited by AE, particularly at higher energies. In this context, a useful metric is also the highest X-ray energy at which Absorption Efficiency reaches 50% ($E_{50\%}$). Notable, the Absorption Efficiency not only directly impacts DE and DQE, but also indirectly influence the detector speed and the maximal counting rate (for photon counting detection mode).

Overall, for comprehensive characterization of a single-readout-channel X-ray detector, we recommend reporting coherently following metrics (Box 1): $E_{50\%}$, DE, NED (in both units: dose and photon number), stating X-ray energy (or X-ray spectrum) at which both DE and NED were measured.

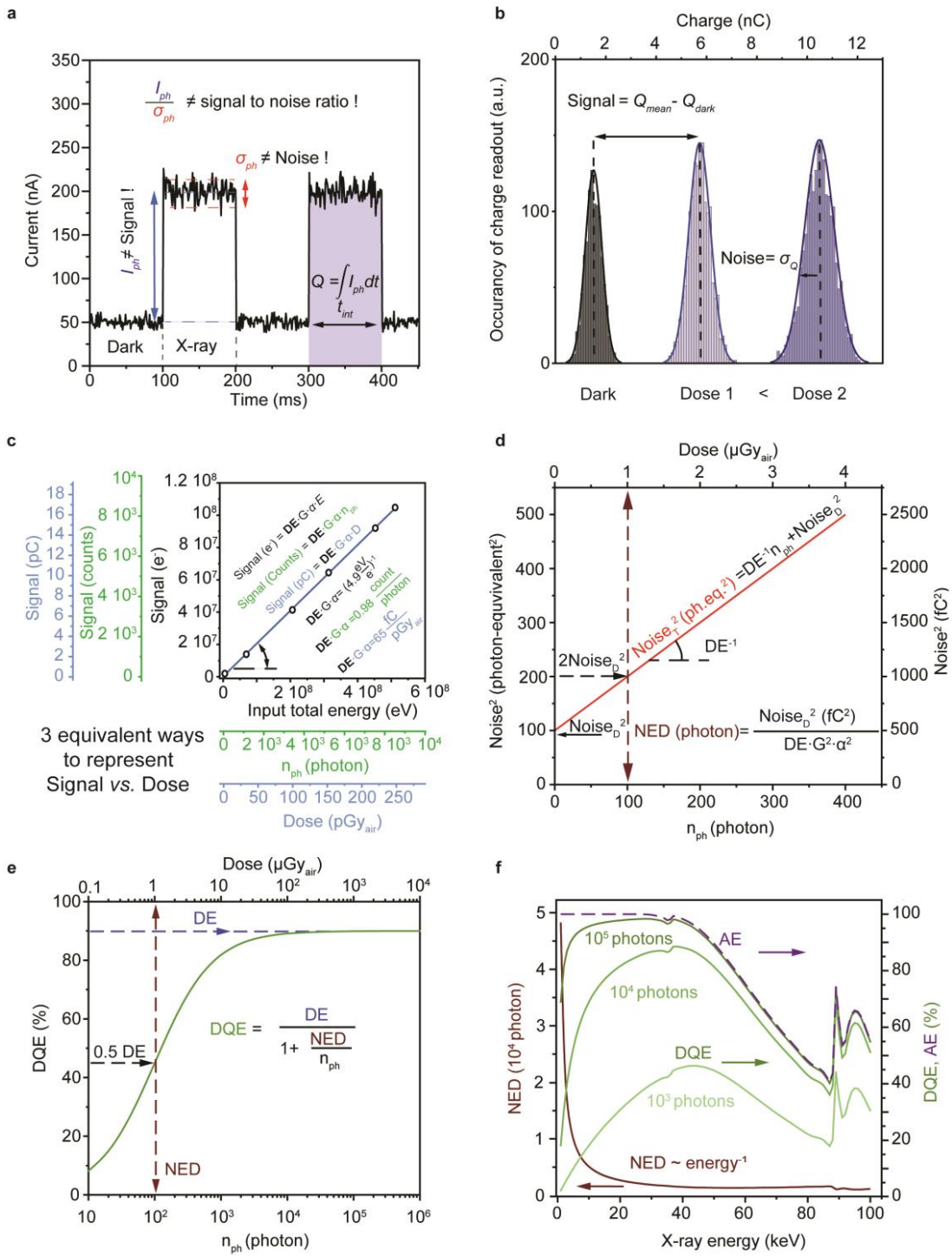

**Fig. 2. Characterizing Single-readout-channel X-ray detector characterization. a**, Current dependence on time in the dark and under X-ray exposure. Wrong yet common definitions of signal and noise are pointed out. The detector integration time and the collected charge under X-ray exposure are indicated. **b**, Distribution of the charge readout occurrence depending on X-ray doses. The detector signal and noise definitions are shown. **c**, Signal amplitude dependence on the radiation dose expressed in 3 equivalent representations in terms of signal and dose units. The signal is expressed in charge units (Coulomb C and electrons e⁻) and counts. Dose units are expressed as total input energy in eV, number of X-ray photons and absorbed dose in Gray in air. **d**, Dependence

of the squared total noise, expressed in photon-equivalent (left Y-axis) and in signal arbitrary units (right Y-axis), on radiation dose. **e,** Dependence of the Detective Quantum Efficiency on radiation dose, based on a model for a single-readout-channel detector, shown as the inset equation. The radiation dose in **d-e** is expressed in Gray in air units (top X-axis) and X-ray photon number (bottom X-axis). **f,** Dependencies of the Noise Equivalent Dose (brown curve, left Y-axis) and Detective Quantum Efficiencies, Absorption Efficiency (green curves and dashed violet curve correspondingly, right Y-axis) on X-ray energy. Detective Quantum Efficiency is shown for different photon numbers (doses).

**Device temporal response characterization**

With the sufficiently fast readout electronics, the temporal response of an X-ray detector is limited by the charge carrier recombination dynamics and the charge transport in a sensor material. Both in direct detection with semiconductors and in scintillator materials, an X-ray photon initially excites an electron from an inner atomic shell to a high level of a conductive band (*via* photoelectric effect or Compton scattering, Fig. 3a). This hot electron then ionizes the sensor material, producing a multitude of high-energy electrons, which then thermalize in the conduction band. Typically, the time-scales of the excitation and multiplication are in the range of femtoseconds and picoseconds, correspondingly, and as such, do not substantially contribute to the temporal characteristics of the detector. On the other hand, decisive are slower (i.e., nanoseconds to seconds), material-specific processes: extraction of charge carriers (for semiconductors) or photoluminescence (for scintillators), along with the carrier trapping and non-radiative recombination (Fig. 3a).

The intrinsic temporal response of the detector (without considering any possible instrumental effects like jitter, time walk *etc.*) can be expressed as the rise and fall times, $\tau_r$ and $\tau_f$, of the detector signal amplitude response between 10% and 90%[32] under a square-shaped X-ray pulse (Fig. 3b,c, case II in Box 1) or a single-photon counting event (case I in Box 1). The steep fast component of the rise or fall time is either (1) the transient time in direct-conversion detectors, the time for collecting the drifting charges, or (2) the luminescence lifetime arising from the radiative recombination in scintillators. The subsequent slower component, if any, usually arises from the initially trapped and subsequently released charges that are then collected at the electrodes or, in scintillator, give raise to the delayed luminescence. Charge trapping may have a complex effect. The signal rise and fall traces become slower for higher trap densities and their greater depth (Fig. 3c). Furthermore, a dependence of the response time on X-ray dose rate might emerge, ultimately yielding faster apparent detector response at high dose rates when the traps are filled (Fig. 3b). It is thus important to measure temporal characteristics at practically relevant doses and dose rates. Here, a typical flaw would be to report, for instance, DE at a low dose rate, while the detector speed is presented for a high dose rate.

The detector time response τ is the slowest value of either of $\tau_r$ or $\tau_f$. The temporal frequency at which DE drops by 3 dB is defined as a cut-off frequency $f_{3dB}$, referred to as detector bandwidth[33]:

$$f_{3dB} = \frac{\ln(9)}{2\pi\tau} \approx \frac{0.35}{\tau}. \quad (13)$$

Notably, besides the DQE roll-off at high-frequencies (Fig. 3d), there exist also the DQE drop at frequencies lower than so-called "1/f noise corner" (typical in sub-Hz range) due to a significant noise rise (Fig. 3d). The frequency range optimal for DQE, *i.e.* working bandwidth $\Delta f$, can thus be defined, which is located between the 1/f noise corner and the $f_{3dB}$ DE cut-off (Fig. 3d, case III in Box 1). Within $\Delta f$, the intrinsic detector noise (Noise$_D$) chiefly originates from the Jonson-Nyquist thermal noise and the dark current electron shot noise, and the noise current $I_n$ is proportional to the square root of the chosen frequency bandwidth $B$[34,35]. Thus, integrated noise charge, *i.e.* Noise$_D$, is proportional to the square root of the integration time[36,37]:

$$\text{Noise}_D = \int_0^{t_{int}} I_n dt \sim \int_0^{t_{int}} \sqrt{B} dt \sim \sqrt{t_{int}}, \quad (14)$$

assuming $B=0.5t_{int}^{-1}$ according to Nyquist-Shannon theorem. Equation (14) yields NED $\propto t_{int}$ (using equation (6)), and, given that the integrated signal $\propto t_{int}$, one obtains SNR $\propto \sqrt{t_{int}}$. The latter scaling is valid for doses that are low enough, *i.e.* comparable to NED, when Noise$_D$ is comparable to photon shot noise (first term in equation (5)).

In the recent literature, a common strive is to report the lowest detectable dose rate, also known as the lowest limit of dose rate detection, LoD, which is a dose rate for achieving SNR$_{LoD}$=3. However, we now show that LoD is, at best, a useless figure of merit since the condition of SNR=3 may be achieved for an arbitrary dose rate if suitably long $t_{int}$ is taken in the experiment (considering SNR$_{out}$ $\propto \sqrt{t_{int}}$). Alternatively, and commonly, LoD is calculated using linear extrapolation of the applied dose rate $Dr$ at the measured SNR$_{out}$ down to SNR$_{LoD}$=3 (assuming SNR $\propto$ dose rate)[37]:

$$\text{LoD} = \frac{\text{SNR}_{LoD}}{\text{SNR}_{out}} Dr = \frac{3}{\text{SNR}_{out}} Dr \sim \left| Dr = \text{const}, \text{SNR}_{out} \sim \sqrt{t_{int}} \right| \sim \frac{1}{\sqrt{t_{int}}}, \quad (15)$$

where one sees that LoD is inversely proportional to $t_{int}$ under the constant dose rate. Thus, we recommend excluding LoD from reporting whatsoever, because the pursuit of the lowest LoD motivates experiments with unreasonably long $t_{int}$ and hence higher acquired dose, showing the absurdity of LoD for the assessment of the materials for low-dose imaging, as illustrated in Fig. 3e. The case (1) and (2) differ in the obtained LoD. Case (1) depicts a low dose rate and longer integration time, yielding lower LoD, but higher accumulated dose compared to case (2). Most importantly, in case (2), employing higher dose rate and much shorter integration time, one receives improvement on all characteristics relevant for low-dose imaging: DQE and SNR are increased, while NED is reduced. Instead of striving for smallest LoD, one has to optimize for lowest NED, whose frequency dependence will determine the optimal $t_{int}$ for a given detector. Specifically, the shortest signal integration time $t_{int}$ to be chosen is determined by the working bandwidth[36], *i.e.* the DQE high-frequency cut-off, discussed above, at which also the NED reaches the minimum (Fig. 3d):

$$t_{int} = \frac{1}{2\Delta f}. \tag{16}$$

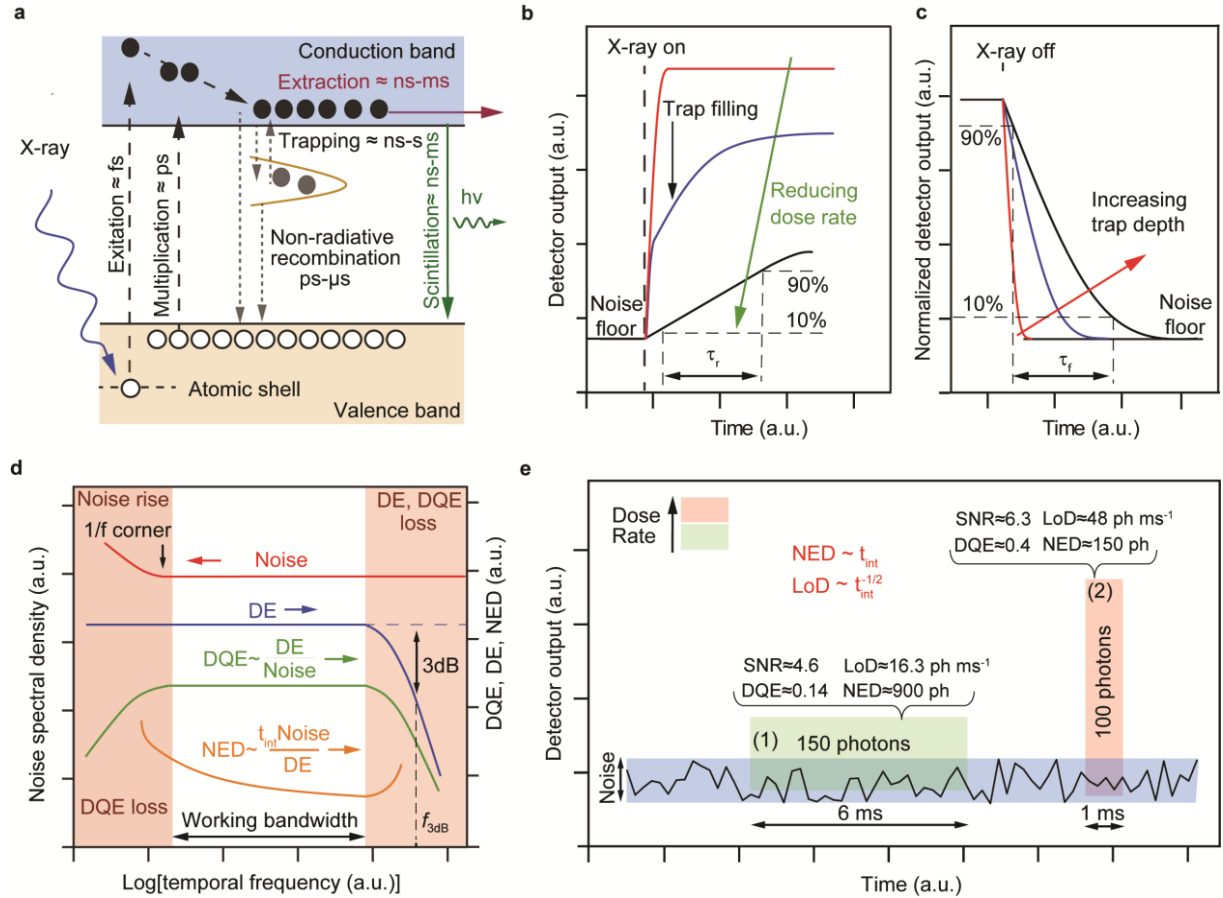

**Fig. 3. Device temporal response characterization. a,** Schematic band structure of direct and indirect conversion semiconductor detector. X-ray charge carriers excitation, multiplication, trapping, non-radiative recombination, scintillation and extraction are illustrated alongside with corresponding time ranges. **b,c,** Effect of charge trapping on the detector response rise time (**b**) and fall time (**c**). **d**, Generic frequency dependencies of the noise spectral density, DE, DQE and NED. **e,** Illustration of the effect of various integration times and X-ray doses on detector performance metrics (DQE, SNR, NED and LoD), emphasizing LoD as a rather irrelevant characteristic for assessing the utility of the detector for low-dose imaging, as better SNR, DQE and NED are obtained with the smaller number of X-ray photons with the detector with improved working bandwidth (lower $t_{int}$).

Overall, for comprehensive characterization of a temporal response from an X-ray detector, we recommend reporting coherently the following metrics (Box 1): $\tau_r$, $\tau_f$, $f_{3dB}$. We encourage the measurements of NED and DE to be taken with the fastest possible integration time, determined by the working bandwidth, which is limited either by the detector or by the application requirements. For example, photon counting detectors must must exhibit response times in sub-ns range in positron emission tomography[38], tens of nanoseconds in computed tomography[39] and up to µs in mammography and radiography[37]. With charge-integration detectors, the response time is set by the X-ray tube pulse duration, which is 3-20 millisecond for modern medical imaging devices[40,41].

**Array detectors**

The characterization described above applies for array detectors as well, albeit the DE and NED are to be reported as a mean value averaged over all channels. The signal and the noise could be determined after a single frame acquisition under homogenous irradiation of the array, as Gaussian distribution described in Fig. 2b can be constructed using data of every individual array readout channel. In addition, the characterization routine for array detectors includes evaluation of the imaging performance, which can be expressed as a DQE dependence on the spatial frequency $f$[19]:

$$\text{DQE}(D,f) = \frac{\bar{S}^2 \text{MTF}^2(f)}{\bar{q} \text{NPS}(f)}, \tag{17}$$

where MTF is a modulation transfer function, NPS is a noise power spectrum, $\bar{S}$ is a mean signal amplitude of all readout channels under homogenous illumination by the mean photon flux $\bar{q}$. The last can be either measured with a reference detector or calculated from equation (4) as:

$$\bar{q} = \frac{n_{ph}}{A} = \frac{D}{\left(\frac{\mu}{\rho}\right)_{air} E_{ph}}, \tag{18}$$

which considers the known dose $D$ and a monoenergetic photon beam. For a polyenergetic irradiation, the averaging over the whole X-ray spectrum for the product $\left(\frac{\mu}{\rho}\right)_{air} E_{ph}$ is required. In equation (17), MTF and NPS are, correspondingly, the dependencies of signal and noise amplitudes on spatial frequency.

MTF can be determined according to the slanted-edge method[15]. From the slanted-edge profile obtained from the X-ray transmittance image of a high-contrast thin slice (Fig. 4a), the edge spread function (ESF(x)) is derived (Fig. 4b) and the MTF is then calculated as:

$$\text{MTF}(f) = F(\text{LSF}(x)) = F\left(\frac{d}{dx}\text{ESF}(x)\right), \tag{19}$$

where $x$ is the position of a corresponding pixel. The line spread function (LSF($x$), Fig. 4c) is the derivative of the ESF($x$); the MTF($f$) is the spatial Fourier transform ($F$) of the LSF($x$) (black curve in Fig. 4d). From the obtained dependence MTF *vs. f,* one can estimate the spatial resolution of the detector system. By definition, MTF($f$) corresponds to the ratio of the image modulation depth, represented by the detector system, to the original modulation depth, *e.g.* transmittance of the object consisting of a sine-wave pattern with the spatial frequency $f$ (inset in Fig. 4d). According to the Rayleigh criterion, two neighbouring points are observed as separate (*i.e.* optically resolved by the detection system) if the intensity signal of the image drops by at least 20% between those points[42]. Thus, the spatial resolution parameter $f_{20}$ is defined as a spatial frequency $f$ at which MTF($f$)=20%. To note, also other intensity-drop threshold criteria for MTF are ubiquitous, within the 10-50% range. For the detector system with pixel pitch $p$ the theoretical MTF limit (red curve in Fig. 4d) is[43]:

$$\text{MTF}_{theory}(f) = \frac{\sin(\pi \cdot f \cdot p)}{\pi \cdot f \cdot p} = \text{sinc}(\pi \cdot f \cdot p). \tag{20}$$

For a 2D array detector, NPS is calculated based on the 2D Fourier spatial transform of its noise image[19]:

$$\text{NPS}(f_x, f_y) = \frac{p_x p_y}{N_x N_y} F(\text{Noise}^2[x, y]) \tag{21}$$

where noise image (Noise[$x,y$]) can be obtained as a root-mean-square deviation from an average difference of $N$ images under homogenous illumination with dose $D$, divided by the correction factor $\sqrt{N}$[44]; $x$ and $y$ are the corresponding pixel coordinates; $f_x$, $f_y$, $p_x$, $p_y$, $N_x$, $N_y$ are, respectively, spatial frequencies, pixel pitch sizes and pixel numbers along denoted directions. When the noise is isotropic for both $x$ and $y$ axis, it can be reduced to a one-dimensional function along one detector matrix direction (black curve in Fig. 4e)[19]:

$$\text{NPS}(f) = \text{NPS}_y(f_y) = \text{NPS}_x(f_x) = \text{NPS}(f_x, f_y = 0). \tag{22}$$

We note that the accurate estimation of the NPS requires multiple averaging over different parts of detector areas, as described in detail in Ref[45]. When the noise is determined by only the Poisson photon statistic, *i.e.* no contribution of the electronic noise, the normalized theoretical NPS (NNPS$_{theory} = \frac{\bar{q}\text{NPS}(f)}{\bar{S}^2}$, red dashed line in Fig. 4e) is equal to 1, *i.e.* for a single-photon counting detector[46]. Once the parameters $\bar{S}$, $\bar{q}$, MTF($f$), NPS($f$) are determined, the DQE *vs.* spatial frequency can be computed (equation (17), black curve in Fig. 4f) in order to evaluate the imaging performance. According to the theoretical model (red curve in Fig. 4f), for doses much higher than NED, DQE follows MTF$^2$($f$)[46].

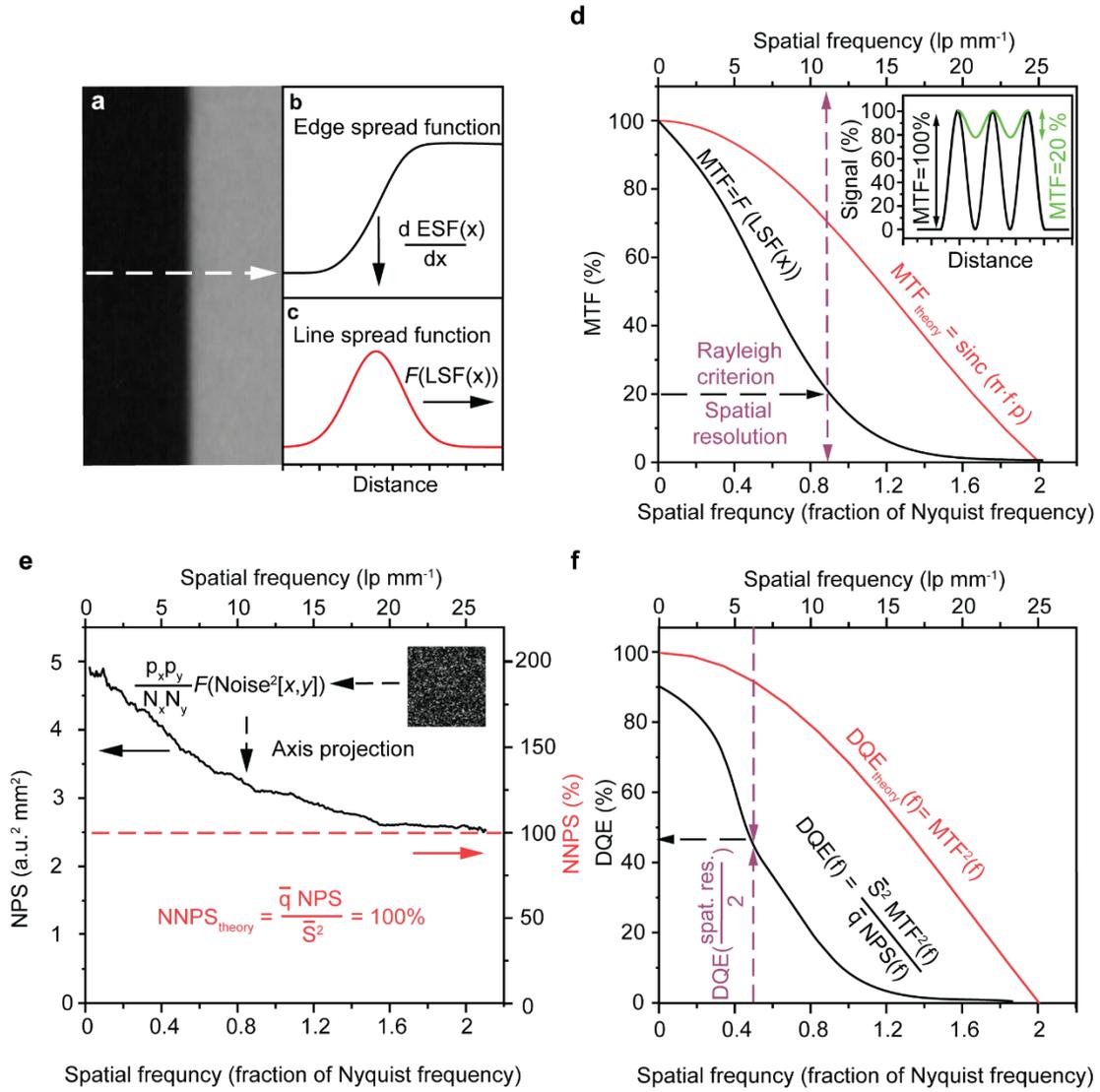

**Fig. 4. Array X-ray detector characterization. a,** X-ray image of the slanged steel edge. **b,** Edge spread function vs. spatial distance, obtained from image **a** (along white arrow). **c,** Line spread function dependence on spatial distance, obtained as a derivative of **b**. **d,** Typical experimental (black curve) and corresponding theoretical (red curve) Modulation Transfer Function dependencies on spatial frequency. Inset shows signal modulation of a sine-shape object, when Modulation Transfer Function equals to 100% (black curve) and 20% (green curve). **e,** Typical experimental Noise Power Spectrum (black solid curve, left axis) and spatial-frequency dependence of the theoretical Normalized Noise Power Spectrum (defined by Poisson photon statistics, red dashed line, right axis). **f,** Typical experimental (black curve) and theoretical (red curve, defined by Poisson photon statistics and pixel pitch) Detective Quantum Efficiency dependencies on spatial frequencies. The spatial frequency in **d**-**f** is expressed in units of line-per-millimeter (top X-axis) and normalized on detector pitch Nyquist frequency (bottom X-axis).

The ratio of the DQE value to its theoretical limit may be used for a detector benchmarking[46]. For more formalized metrics, we propose a cumulative figure of merit - the detector quality factor or DQE index (DQEi) - as a ratio of the obtained DQE to theoretical DQE at the spatial frequency of the half of the spatial resolution and the dose that equals the largest value from either of 100 NED or 100 photons:

$$\text{DQEi} = \frac{\text{DQE}(f = 0.5\ f_{20}; D = \max[100 \times \text{NED}, 100 \times \text{photons}])}{\text{DQE}_{\text{theory}}(f = 0.5\ f_{20})}. \tag{23}$$

We hope that the presented characterization framework will aid in objective evaluation and comparison of novel X-ray detector materials, particularly when the research is motivated by the eventual utility in medical imaging. The ready-to-use calculator of these characteristics is provided in several formats: as a Mathcad worksheet (Supplementary File), MATLAB application (Supplementary File) and as a website[20]. The presented formalism applies to both direct conversion and scintillation detectors. Table 2 enlists all characteristics, recommended for reporting, while Box 1 proposes the respective characterization checklist.

**Table 2. X-ray detectors figures of merit.**

| Figure of Merit | Estimated from | What is characterized |
|---|---|---|
| Noise equivalent dose | Noise *vs.* dose | Low dose performance |
| Detection efficiency | Signal *vs.* dose, Noise *vs.* dose | Fraction of detected photons |
| X-ray energy at 50% absorption efficiency | Absorption efficiency *vs.* X-ray energy | Applicable X-ray energy range |
| Rise and fall times or cut-off frequency | Signal *vs.* time | Detector speed |
| Spatial resolution | Modulation transfer function *vs.* spatial frequency | The smallest resolvable spatial feature |
| Detective Quantum Efficiency index | Experimental and theoretical Detective Quantum Efficiency dependencies on spatial frequency | The detector quality factor |

**Box 1. Characterization checklist.**

**Single-readout channel detectors:**

**Noise equivalent dose**
1. Determine the slope from the signal *vs.* dose dependence (DE·$G$·$\alpha$ product at Fig. 2c).
2. Using the offset and the slope from the noise *vs.* dose dependence (Fig. 2d) and determined before DE·$G$·$\alpha$ product, estimate NED both in dose and photon number units.
3. Using Eq. 10, plot NED dependence on X-ray energy (Fig. 2f).

**Detection efficiency**
*Case A - photon-counting detection mode.*
Determine directly from the slope from the signal *vs.* dose dependence (DE·$G$·$\alpha$ product at Fig. 2c), as in this case $G$·$\alpha$ = 1 count/photon.
*Case B – charge-integration mode, when $G$·$\alpha$ product is known.*
Determine directly from the slope from the signal *vs.* dose dependence (DE·$G$·$\alpha$ product at Fig. 2c).
*Case C – charge-integration mode, when $G$·$\alpha$ isn't known.*
1. Determine the slope from the signal *vs.* dose dependence (DE·$G$·$\alpha$ product at Fig. 2c).
2. Measure the noise *vs.* dose dependence (Fig. 2d). Using DE·$G$·$\alpha$ product to express noise in photon-equivalent units. Determine DE from the fit by Eq. 7. of the noise vs. dose dependence.
*Case D – quick evaluation for any detection mode.*
The upper limit of detection efficiency can be determined as equal to the absorption efficiency (Eq. 9.)

**X-ray energy at 50% absorption efficiency**
Determine from X-ray absorption efficiency dependence on X-energy (Fig. 2f).

**Detective Quantum Efficiency dependence on dose and X-ray energy**
Using previously determined DE, NED and X-ray absorption efficiency, plot DQE *vs.* dose and energy (Fig. 2e,f). The latter is applicable only using (quasi)monoergentic X-ray spectrum. Otherwise, report the used X-ray beam quality only.

**Rise and fall time, cut-off frequency**
*Case I - photon-counting detection mode.*
Determine the detector speed from the rise time with single-photon counting event. Calculate cut-off frequency using Eq. 13.
*Case II - charge-integration mode, when X-ray is possible to apply as square pulse.*
Determine the detector speed from the slowest of the rise time and the fall time (Fig. 3b). Calculate cut-off frequency using Eq. 13.
*Case III - charge-integration mode, when variation of X-ray modulation frequency is possible*
Determine cut-off 3dB frequency the signal dependence on frequency (Fig. 3c), calculate response time through Eq. 13.

**Array detectors** (in addition to the above reported metrics with statistical distribution among pixels)**:**

**Spatial resolution**
Determine the spatial resolution from Modulation transfer function dependence on spatial frequency (Fig. 4a, b).

**Noise power spectrum A**
Using noise data, plot NPS *vs.* spatial frequency.

**Detective Quantum Efficiency dependence on spatial frequency**
Using MTF, NPS *vs.* dependencies and Eq. 17, plot DQE *vs.* spatial frequency.

**Detective Quantum Efficiency index**
Using all above metrics, determine DQEi according to Eq. 23.

**List of abbreviations in the order of their appearance in the main text**

DQE – detective quantum efficiency

NED – noise equivalent dose

DE – detection efficiency

Detective Quantum Efficiency index – DQEi

$D$ – radiation dose, incoming to the detector volume

$\alpha$ – conversion coefficient, which determines the physical transformation of the dose (or X-ray photon number) into the charge (*i.e.,* X-ray sensitivity) or optical photons (*i.e.,* light yield)

$G$ – electronic gain of the detector (sometimes includes a photoconductive gain of active materials).

$e$ – elementary charge

$W_{\pm}$ – electron-hole pair creation energy in a detector material by an X-ray ionization

$\left(\frac{\mu}{\rho}\right)_{air}$ – X-ray mass energy absorption coefficient in the air (X-ray-energy dependent)

$E_{ph}$ – X-ray photon energy

$A$ – detector area

$\text{Noise}_T$ – total detector noise

$\text{Noise}_D$ – intrinsic detector noise

$n_{ph}$ – X-ray photon number incident to the detector volume

$\text{SNR}_{out}$ – signal-to-noise ratio experimentally measured by detector

$\text{SNR}_{in}$ – the highest limit for a signal-to-noise ratio for a given dose, defined by photon shot noise

$E_{mes}$ – X-ray photon energy, which was used to measure DE and NED

$\mu$ – linear X-ray absorption coefficient (dependent on $E$, which can be calculated using atomic content)

$d$ – the detector thickness

$\tau_r$ – the detector signal rise time on response of square shape X-ray pulse

$\tau_f$ – the detector signal fall time on response of square shape X-ray pulse

$f_{3dB}$ – temporal frequency at which DE decreases by 3 dB, comparing to near 0 Hz value.

$\Delta f$ – working bandwidth, *i.e.* frequency range for stable DE and DQE

LoD – lowest detection limit of a dose rate

Dr – dose rate

$f$ – spatial frequency

MTF – modulation transfer function

NPS – noise power spectrum

$\bar{S}$ – mean signal amplitude of all readout channels under homogenous illumination of mean photon flux

$\bar{q}$ – mean X-ray photon flux

ESF($x$) – edge spread function

LSF($x$) – line spread function, which is the derivative of the ESF($x$)

$F$ – spatial Fourier transform

$p$ – pixel pitch

$\text{NNSP}_{theory}$ – normalized noise power spectrum, for theoretical case, when only the Poisson photon statistic defines noise (no contribution of electronic noise)

**Acknowledgments**
This work was financially supported by the Swiss Innovation Agency (Innosuisse) under grant agreement 46894.1 IP-ENG and by ETH Zürich through the ETH+ Project SynMatLab: Laboratory for Multiscale Materials Synthesis. J. H. acknowledges the financial support from the National Institutes of Health under award 1R01EB033439. The Authors thank Andriy Lomako (Teledyne DALSA, Healthcare X-ray Solutions) as well as Michal Bochenek and Roger Steadman (ams-OSRAM) for insightful discussions.

**Additional Information**

**Materials & Correspondence** Maksym V. Kovalenko (mvkovalenko@ethz.ch)

**Correspondence and requests for materials** should be addressed to M.V.K.

**Data availability**
*Mathcad worksheet and MATLAB application with step-by-step calculations of the protocol described in this Perspective on examples of typical X-ray detectors materials (CdTe, Si, a-Se, GOS, MAPbI$_3$) are available as Supplementary Data, a non-proprietary online version of the calculation tool is available online at the link in Ref.20. The tutorial videos, describing data processing and calculation processes are available at the KovalenkoLab YouTube channel under following links:*

*Mathcad worksheet tutorial:* https://youtu.be/b1Fb-BmKCtY
*MATLAB software tutorial:* https://youtu.be/RGa9SugcS18
*Website tutorial:* https://youtu.be/r9I82cdtKuU

**Code availability**
*MATLAB application codes are available in public GitHub repository:* https://github.com/VitaliiBartoshETH/perovXimager.git .